\documentclass{article}
\usepackage{spconf, amsmath, graphicx}
\usepackage{indentfirst}
\usepackage{float}
\usepackage{graphicx}
\usepackage{subcaption}
\usepackage{adjustbox}
\usepackage{siunitx}
\usepackage{color}
\usepackage{amsfonts}
\usepackage{booktabs}
\usepackage{algorithm}
\usepackage[noend]{algpseudocode}
\usepackage[normalem]{ulem}
\useunder{\uline}{\ul}{}
\usepackage{caption}
\title{lCuts: Linear clustering of bacteria using recursive graph cuts}
\name{J. Wang $^{\dagger}$, T. Batabyal $^{\dagger}$, M. Zhang $^{\ddagger}$, J. Zhang $^{\ddagger}$, A. Aziz $^{\ddagger}$, A. Gahlmann $^{\ddagger}$ and S. T. Acton $^{\dagger}$}
\address{$^{\dagger}$Department of Electrical \& Computer Engineering and $^{\ddagger}$Department of Chemistry\\
University of Virginia, Charlottesville, VA 22904, USA}
\begin{document}
%\ninept
%
\maketitle
\begin{abstract}
Bacterial biofilm segmentation poses significant challenges due to lack of apparent structure, poor imaging resolution, limited contrast between conterminous cells and high density of cells that overlap. Although there exist bacterial segmentation algorithms in the existing art, they fail to delineate cells in dense biofilms, especially in 3D imaging scenarios in which the cells are growing and subdividing in a complex manner. A graph-based data clustering method, $\mathcal{L}$Cuts, is presented with the application on bacterial cell segmentation. By constructing a weighted graph with node features in locations and principal orientations, the proposed method can automatically classify and detect differently oriented aggregations of linear structures (represent by bacteria in the application). The method assists in the assessment of several facets, such as bacterium tracking, cluster growth, and mapping of migration patterns of bacterial biofilms. Quantitative and qualitative measures for 2D data demonstrate the superiority of proposed method over the state of the art. Preliminary 3D results exhibit reliable classification of the cells with $97\%$ accuracy.

\end{abstract}
\section{Introduction}
\label{sec:intro}
\vspace{-0.2cm}
Analyzing cellular behavior of individual bacteria in a biofilm is a key for biologists and biochemists to understand biofilm growth, in diverse applications such as electrical power and public health research \cite{nadell2016spatial}. Lack of knowledge in macroscopic biofilm properties (e.g. size, shape, cohesion / adhesion) that emerge from the behaviors of individual bacteria in different micro-environment is a major barrier in biofilm studies. To make up for the deficiency, an advanced image analysis toolkit for segmenting individual cells is in high demand along with efficient image acquisition methods, such as using super-resolution technology \cite{gahlmann2014exploring}\cite{sahl2017fluorescence} that overcome the diffraction limit of traditional optical microscopy techniques. 

\begin{figure}
	\centering
	\includegraphics[width=\columnwidth]{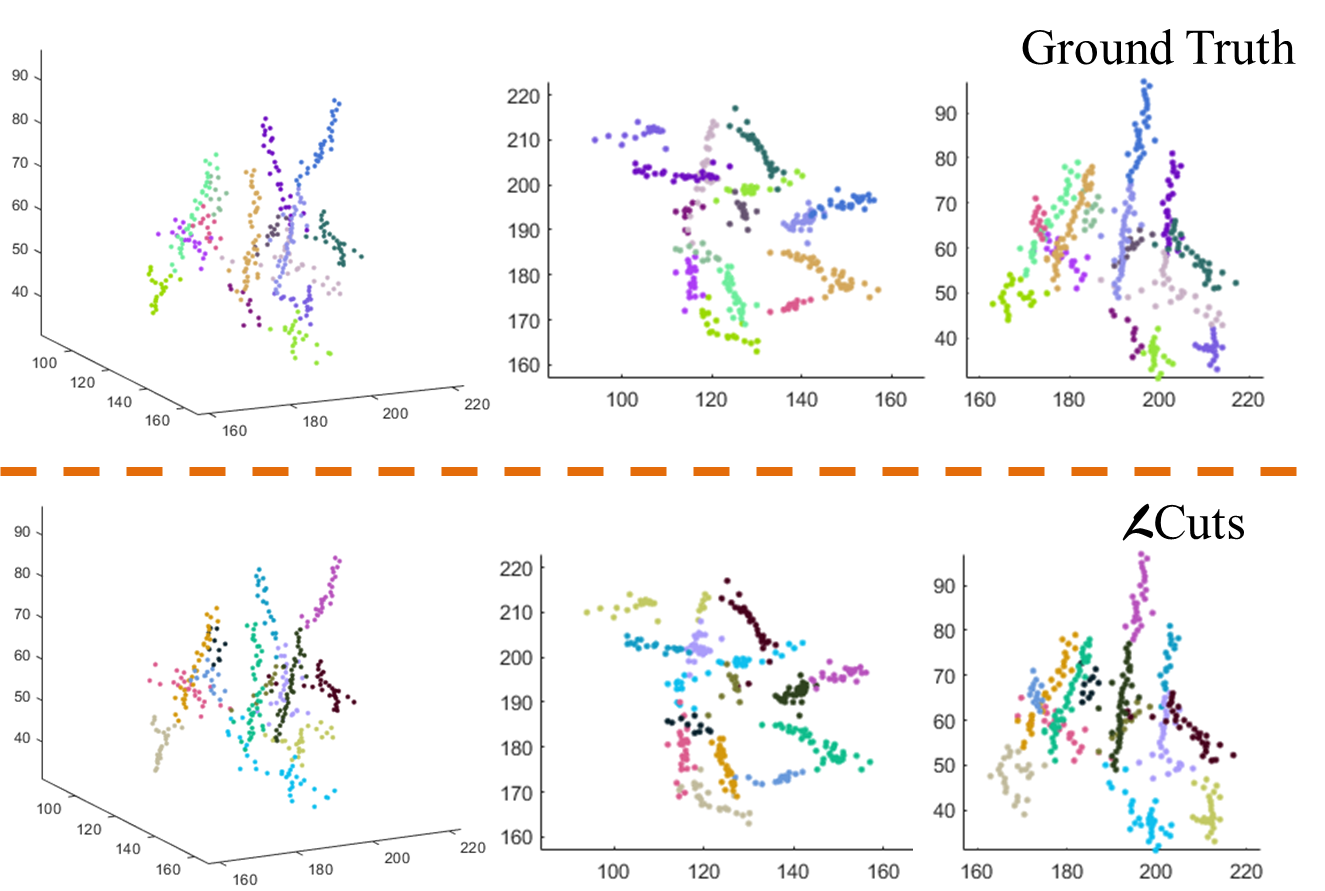}
	\vspace{-0.5cm}
	\caption{\small{Performance of $\mathcal{L}$Cuts on 3D point cloud data comparing with manually grouped ground truth. The counting accuracy is $97\%$ and grouping accuracy is $90\%$ . Three viewpoints from left to right: 3D view, $xy$-plane and $yz$-plane. }}
	\label{fig:3dres}
	\vspace{-0.8cm}
\end{figure}

The segmentation of individual bacterial cells in dense bacterial biofilms is a challenging problem. One of the major challenges to the state of the art comes from the presence of inhomogeneous fluorescence intensity within a single cell or across multiple cells. When using standard level set segmentation methods \cite{acton2009biomedical} and level sets using Legendre polynomials as basis functions \cite{mukherjee2015region}, the segmentation fails where the contrast between the cells and background is weak. The watershed algorithm \cite{vincent1991watersheds}\cite{wright1997watershed} uses the gradient flow to identify the morphological changes along segment contours, whereas \cite{reyer2017automated} \cite{he2015icut} separate large segments at the concavities. With both approaches, situations where the intensity of the regions of interest is non-homogeneous often lead to segmentation errors. Other edge-based parametric methods \cite{ray2002active}\cite{mansouri2004constraining}are insufficient given the subtle and often absent boundaries between cells that are densely packed in three dimensions. 

To achieve 3D cell segmentation, the authors in \cite{sadanandan2016segmentation} presented a technique to track bacteria in a dense mono-layer by way of 3D time-lapse images. This solution employs an iterative threshold-based approach, which is heavily dependent on high contrast between the signal and background in the images. Yan \textit{et al}. \cite{yan2016vibrio} proposed a single cell tracking toolkit based on marker controlled watershed and threshold techniques. This method allows tracking of bacterial growth in multi-layered biofilms when florescence intensity is uniform and void spaces between cells are readily discernable, but struggles with the detection of individual bacteria when cells are closely packed or when inter- and intra-cellular fluorescence intensity are not heterogeneous. Building on the work in \cite{yan2016vibrio}, Hartmann \textit{et al}. \cite{hartmann2018emergence} recently reported a solution to 3D segmentation in confocal images of biofilms that exploits prior knowledge of cell size to segment low density biofilms. As this method, like that of \cite{yan2016vibrio}, is watershed-based, it suffers from similar drawbacks. In \cite{wang2017bact}, the authors attempted to solve the problem via constructing single-cell regions to ensure the gap between neighboring cells in a seeded iterative active contour approach. The single cell identification performance degrades in the cases where the contrast between cells and voids in the biofilm is low.

As a solution to overcome the aforementioned limitations, namely the difficulty in segmenting dense aggregations in large biofilm with non-homogeneous inter- and intra-cell intensities, a novel approach is proposed in this paper with two major contributions: 
\vspace{-0.2cm}
\begin{itemize}
\item The bacterial cell segmentation problem is transformed into a data clustering problem by generating pointillist data that represents the regions of interest; 
\vspace{-0.2cm}\item A recursive multi-class linear data clustering algorithm ($\mathcal{L}$Cuts) is proposed that is capable of finding the linear structures in point cloud data where cell boundaries may be ambiguous.
\vspace{-0.2cm}
\end{itemize}
Our approach is built on the following insight:  Even though the raw image data does not show distinct boundaries in intensity between densely packed cells, we are still able to reliably compute local intensity maxima that delineate the central axis of each cell. Therefore, the proposed $\mathcal{L}$Cuts algorithm first derives these maximal points and then partitions them based on the approximate co-linearity of points. Moreover, this local maximum-based initialization translates seamlessly and robustly into the 3D imaging and 3D segmentation problem. 
\vspace{-0.2cm}
\section{Linear Clustering Algorithm}
\label{sec:algorithm}
\vspace{-0.2cm}
Numerous algorithms exist in the clustering community that group the data by finding the similarities between classes. Distance, number of neighbors, density and predefined probability distribution functions are the major perspectives for measuring similarities between points in the trending literature, such as k-means \cite{macqueen1967some}, DBSCAN \cite{ester1996density}, DensityClust \cite{rodriguez2014clustering}. Among those, density based clustering methods (\cite{ester1996density,rodriguez2014clustering}) detect non-spherical arbitrary clusters; however, they are still limited in precisely classifying linear groups as discussed in the comparison in \textbf{sec \ref{sec:res}}.
The Hough transform \cite{mukhopadhyay2015survey} is well known in detecting lines in the space, but the approach is not sufficient for delineating cells that are intersecting and is also computationally expensive. Unlike k-means and DensityClust, our approach does not require manual intervention in order to locate appropriate number of clusters. Incorporation of structural constraints, such as the distance limit and the eccentricity of the bacteria into $\mathcal{L}$Cuts obviates the need for \textit{a priori} information regarding the number of clusters (in our case, bacteria), making $\mathcal{L}$Cuts a fully automatic approach. 

\begin{figure}
	\centering
	\includegraphics[width=0.9\columnwidth]{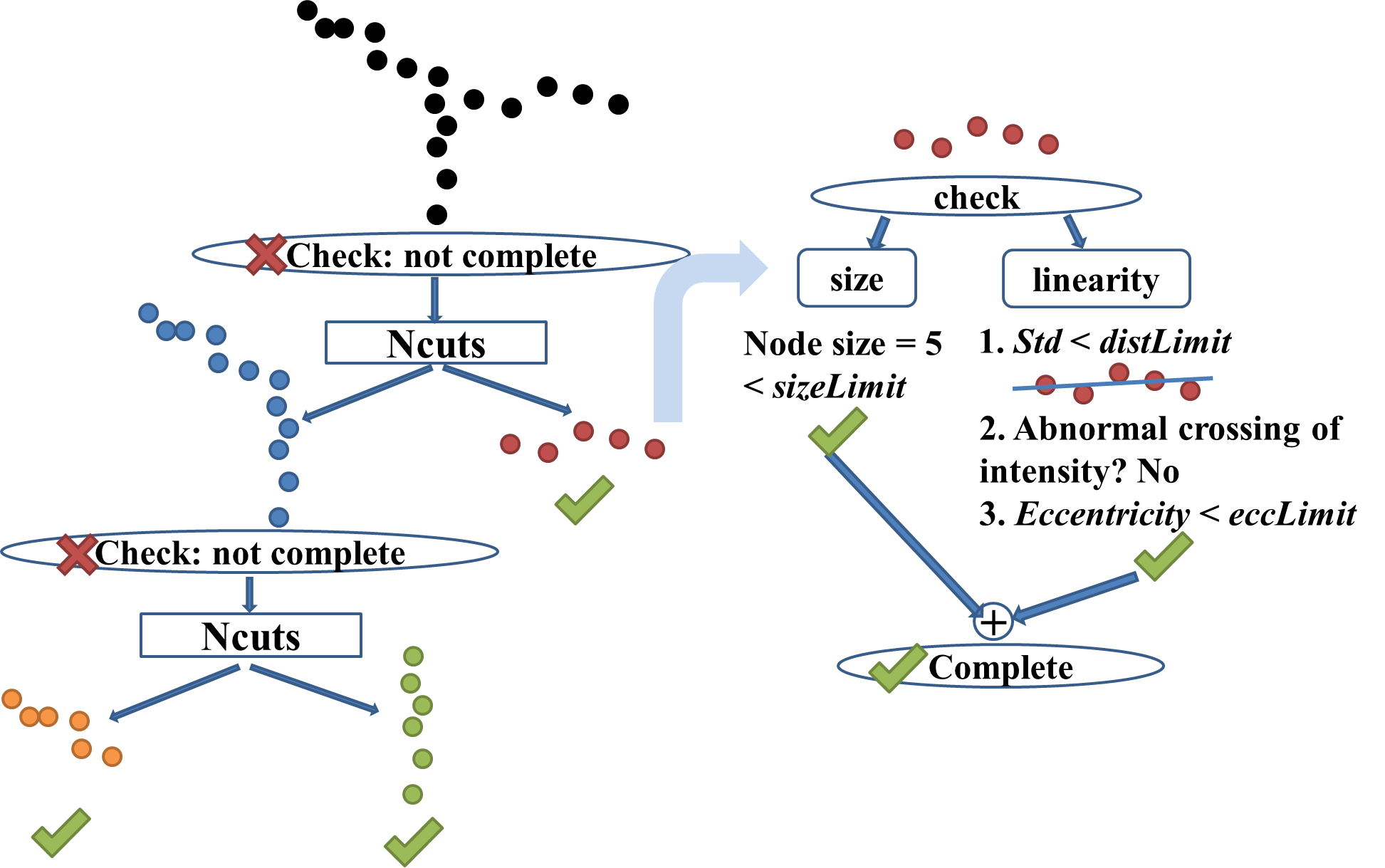}
    \vspace{-0.45cm}
	\caption{\small{Intuitive work flow for the recursive program. Left: an example of bi-partition decision tree. Right: detailed example for checking the stopping criterion of component red. Here, $sizeLimit$, $distLimit$ and $eccLimit$ are parameters based on prior biological information.}}
	\label{fig:checksc}
	\vspace{-0.6cm}
\end{figure}

In the paper, we propose a recursive graph cuts algorithm (see work flow in \textbf{Fig. \ref{fig:checksc}}) for efficient computation to find the linear groups in the point cloud data. The algorithm can be primarily divided into three parts: construct the graph (\textbf{sec \ref{ssec:graph}}), compute the bi-partition solution, and recursively re-partition until the stopping criterion is satisfied (\textbf{sec \ref{ssec:sc}}). The bi-partition solution to separate the nodes (the local maxima) is inspired by \cite{shi2000normalized}. They addressed the problem "how to find the precise groups in an image" as normalized graph partitioning, where an image is partitioned into two groups, $A$ and $B$, by disconnecting the edges between these two groups.
\vspace{-0.3cm}

\subsection{Graph construction}
\label{ssec:graph}
\begin{figure}[H]
\vspace{-0.5cm}
	\centering
	\includegraphics[width=0.9\columnwidth]{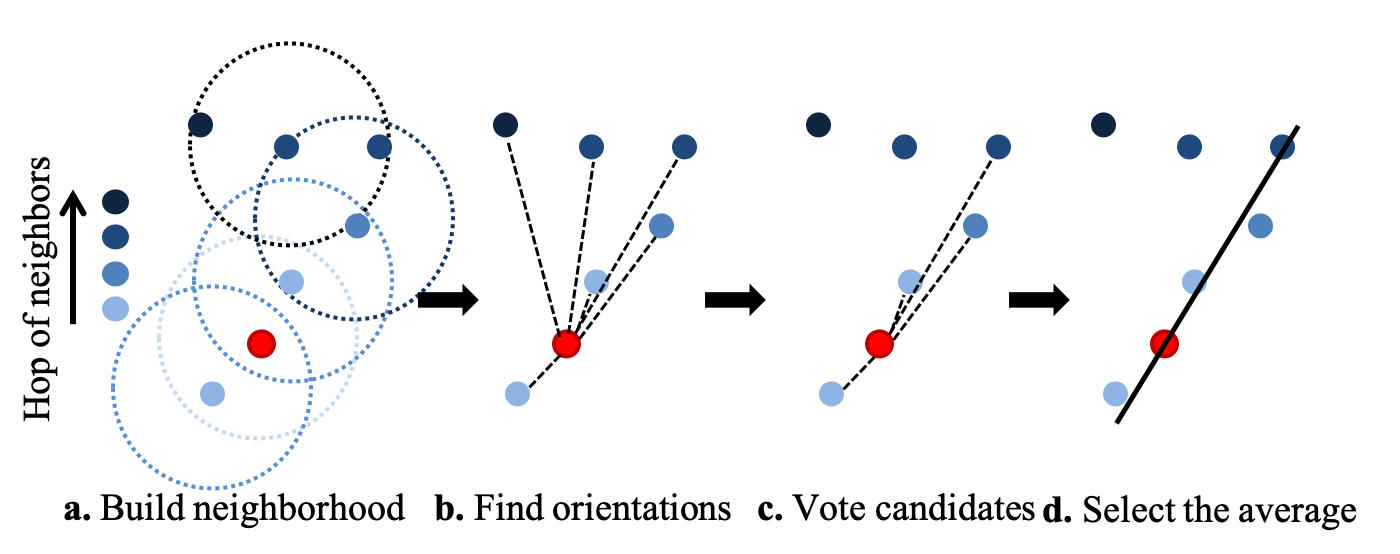}
    \vspace{-0.3cm}
	\caption{\small{An illustration of majority voting. (a) A 4-hop "neighborhood" example. Each hop-neighbor is found within a specified distance (dashed circles) to node. (b) The dashed lines connecting target node with all the other nodes in the neighborhood are possible orientations. (c) Those orientations that have larger relative angles with respect to the orientations are excluded from the candidates. (d) The direction to represent the target node is determined as the average orientation from the candidates.}}
	%Comment: "is defined if nodes are connected to the target node within certain levels." unclear to me
	\label{fig:findDir}
	\vspace{-0.5cm}
\end{figure}

\vspace{-0.2cm}
\textbf{Nodes:} The nodes (local maxima along the ridgeline of a cell) in the constructed graph have two features: location ($nodeLoc$) and direction ($nodeDir$). Location is simply the Cartesian position of the node. Direction of each node is the principal axis direction of the ridgeline computed via majority voting (\textit{see} \textbf{Fig. \ref{fig:findDir}}). A "neighborhood" consists of multi-hop neighbors that is constructed for voting. In graph theory, a "hop" between two nodes is defined as the number of edges that one has to traverse in order to reach from one node to the other node.
%A multi-level "neighborhood" is defined by including the closest nodes to the target node if they are connected within certain number of intermediate nodes (so called neighbors in "levels") and within user defined maximum radius to these intermedium (dashed circles in \textbf{Fig. \ref{fig:findDir}a}). The possible orientations include all the connections from the target node to all the nodes in the previously defined "neighborhood." (\textbf{Fig. \ref{fig:findDir}b})) To increase the accuracy of the voted direction, one can enrich the possible orientations by adding the connections from the first level neighbors to the others as well. Then, a 

% Question: would the caption clear enough to express the paragraph above?
A $N_r\times N_p$ accumulator is set up for the majority voting. One dimension of this accumulator represents the $N_p$ possible orientations ($\mathbf{p}$) in $N_p$ bins (see \textbf{Fig. \ref{fig:findDir}b}). Another dimension corresponds to the quantized relative angles ($\phi$) with $N_r$ bins, where $\phi$ is computed from each possible orientation to all the others. Here, $N_r$ is chosen based on the "hop" number. The accumulator will count the number of parameter pairs ($\mathbf{p},\phi$) that lie in each bin. Within the first bin of $\phi$, the orientations with the largest value are selected which give the candidate directions. These candidates are averaged to yield the major direction for the target node.
%Comment: "each level of the neighboring nodes" not clear

\textbf{Adjacency matrix:} The adjacency matrix reflects the likelihood if two nodes are in the same group. Suppose there are $N$ nodes in the graph, then the dimension of the adjacency matrix is $N \times N$. Each attribute in the matrix represents the connectivity and edge weight between two nodes ($i$, $j$), which measures the similarity of their features according to:
\vspace{-0.1cm}
\begin{equation}
\label{eq:adj}
w_{ij} =  w_{distance}\cdot w_{direction} \cdot  w_{intensity}
\vspace{-0.2cm}
\end{equation}
Three similarity measures are involved: the Euclidean distance of locations of nodes (\textbf{eq \ref{eq:adjdist}}), the relative angle between major directions (\textbf{eq \ref{eq:adjdir}}) and the dissimilarity of intensity along the segment connecting two nodes (\textbf{eq \ref{eq:adjint}}). 

The first term is straightforward with an additional condition that sets the weights to be zero when two nodes are farther than a given distance $r$ (set by maximum cell length). 
\vspace{-0.2cm}
\begin{equation}
\label{eq:adjdist}
w_{distance} =
 \begin{cases}
   e^{- D_{ij} ^2/\sigma_D^2}, & {\text{if $ D_{ij} \leq r$}}\\
    0, & {\text{otherwise}}
  \end{cases}
  \vspace{-0.2cm}
\end{equation}
where $ D_{ij} = ||\text{nodeLoc}_i-\text{nodeLoc}_j||_2$ and $\sigma_D$ reflects the allowed variance for distance between nodes. The second part measures the angle difference between two node directions, called relative angle. Given two node directions, the relative angle ($\theta$) is the cosine term that varies from 1 to 0 as $\theta$ becomes larger. Then the corresponding weighting is given by:
\vspace{-0.1cm}
\begin{equation}
\label{eq:adjdir}
w_{direction} =  e^{-(\cos(\theta)-1)^2/\sigma_T^2}
\vspace{-0.2cm}
\end{equation}
By adjusting $\sigma_T$ , one can control the variance of relative angles within each group. 

\begin{figure}
\vspace{-0.2cm}
	\centering
	\includegraphics[width=0.8\columnwidth]{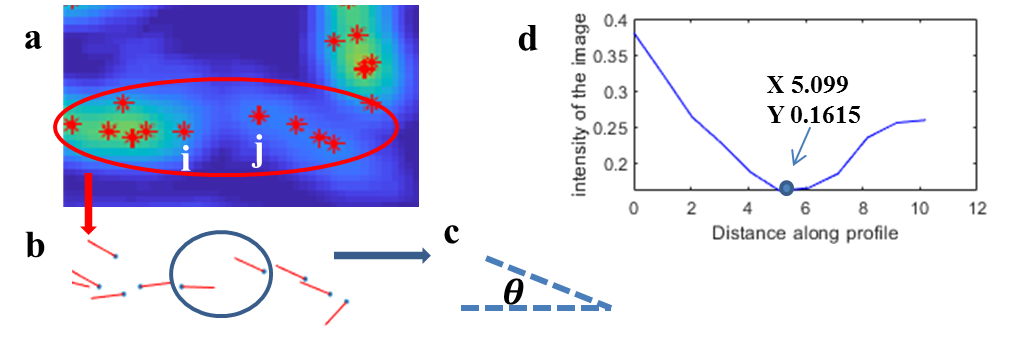}
    \vspace{-0.3cm}
	\caption{\small{Illustration and motivation of defining intensity on edge weight. \textbf{a}: Nodes are denoted as red asterisk. \textbf{b}: After extracting the node directions from the red region in \textbf{a}, it is still hard to separate two groups as the relative angle (in \textbf{c}) and relative distance (shown in d, the distance is 10) are close. In this case, we evaluate the intensity along the connection of the two nodes. The intensity changes are shown in \textbf{d}. The intensity weighting is then assigned as the lowest intensity lower than $thresh$.}}
	\label{fig:adjint}
	\vspace{-0.65cm}
\end{figure}

The third term in (1) detects the intensity dissimilarity along the segment joining two nodes in the image, which is defined as:
\vspace{-0.3cm}
\begin{equation}
\label{eq:adjint}
w_{intensity} =  
\begin{cases}
 \min I_{i\rightarrow j}, & {\text{if $\min I_{i\rightarrow j}\leq \textit{thresh}$}}\\
    1, & {\text{otherwise}}
  \end{cases}
  \vspace{-0.2cm}
\end{equation}
Here, \textit{thresh} equals the difference between the midrange (\textit{Mid}) of all the nodes and the variance (\textit{Var}) of the constituent node intensities. In the case that the nodes have no intensity information, this term can be set as $\mathbf{1}$. Otherwise, we extract the intensities along the connecting segment from node $i$  to node $j$ from the image as shown in \textbf{Fig. \ref{fig:adjint}} and compute the lowest intensity along the segment and compare to \textit{thresh}. 

\vspace{-0.2cm}
\subsection{Stopping criterion for recursion}
\label{ssec:sc}
\vspace{-0.2cm}
Two stopping conditions are checked after each bi-partition level to decide the completeness of the recursion.

\textbf{Criterion 1 - size:} The preliminary components that are less than $sizeLimit$ have the potential to be an individual group. This $sizeLimit$ is a user defined parameter. For the application we discuss in the paper, we used the prior bio-information of the maximum length of the bacterium to determine the value.

\textbf{Criterion 2 - linearity:} This criterion is designed for preserving the linear groups with different size from the potentials (after criterion 1). Intuitively, if a single component is found (see black nodes in \textbf{Fig. \ref{fig:checksc}}) and it is less than the maximum size limit as specified, it may not be a finalized group as linearity remains to be checked. Three aspects are checked to ensure the linearity: 
(1) \textit{Standard deviation} (\textbf{Std}) from nodes in the group to the least square fitted line; (2) \textit{Intensity changes} between the nodes within the group (as explained in \textbf{Sec \ref{ssec:graph}}); 
(3) \textit{Eccentricity} of the group. This is an optional condition based on the data type. For linear components, the eccentricity ($eccLimit$) is closer to 1; while, for circular components, it is closer to 0.
\vspace{-0.2cm}
\section{Application and Analysis}
\label{sec:res}
\vspace{-0.2cm}
\subsection{Experiments on bacterial images}
\begin{figure}[H]
\vspace{-0.3cm}
	\centering
	\includegraphics[width=\columnwidth, height = 0.38\columnwidth]{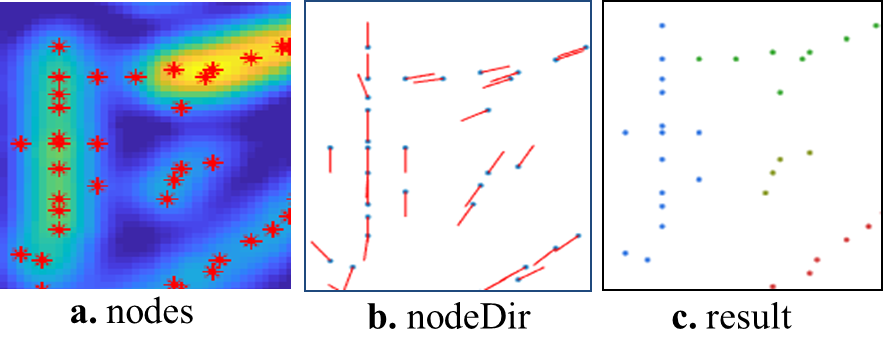}
    \vspace{-0.5cm}
	\caption{\small{An example performance of $\mathcal{L}$Cuts with constructed graph features. (a) Nodes are marked in red asterisks. (b) Red lines show the major direction features for each nodes (blue dots). (c) $\mathcal{L}$Cut clustering results for the constructed graph. }}
	\label{fig:AppExample}
	\vspace{-0.3cm}
\end{figure}

\vspace{-0.2cm}
For qualitative and quantitative assessments, $\mathcal{L}$Cuts is tested on 10 two-dimensional point cloud data which are generated from bacterial images using Airyscan microscopy. From these images, we obtain prior information regarding the longest bacterium in the dataset (approximately maximum 60 pixels in length and 15 pixels in width, where each pixel is $46nm \times 46nm $). The typical data have 250 to 600 nodes with approximately 20 to 60 cells observed.
\begin{figure}[H]
	\centering
	\vspace{-0.35cm}
	\includegraphics[width=\columnwidth]{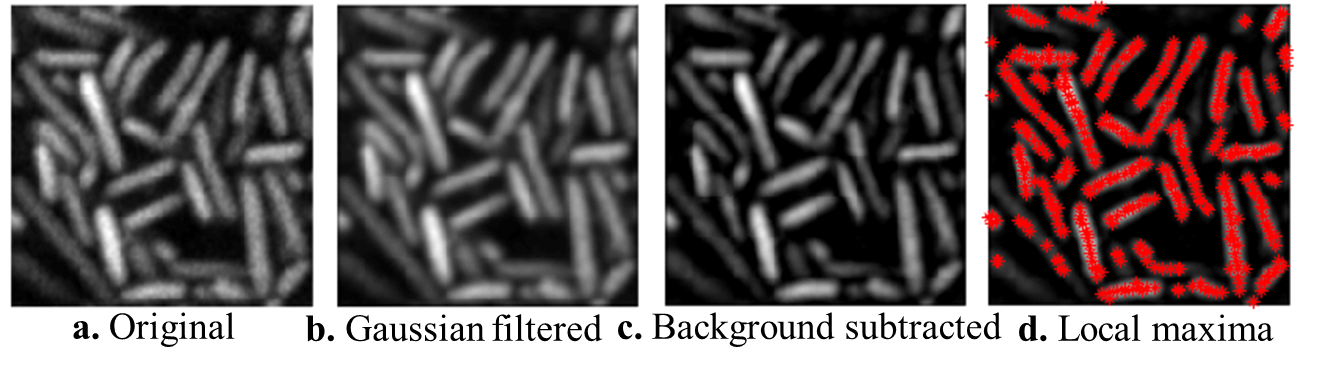}
    \vspace{-0.8cm}
	\caption{\small{Pipeline for finding nodes from bacterial images. Step 1: Filter the original image with a Gaussian kernel (a $\rightarrow$ b). Step 2: Enhance the signals in the image via background subtraction (b $\rightarrow$ c). Step 3: Find the local maxima (c $\rightarrow$ d). Step 4: Clear the points if they have no neighbors or overlap with other points and rest are the found nodes (red asterisks in d).}}
	\label{fig:findnodes}
	\vspace{-0.5cm}
\end{figure}
To build the graph, we generate the point cloud data following the pipeline in \textbf{Fig. \ref{fig:findnodes}}. An experimental result and corresponding node features is shown in \textbf{Fig. \ref{fig:AppExample}}.

\subsection{Qualitative and quantitative comparison}
\begin{figure}
	\centering
    \vspace{-0.3cm}
	\includegraphics[width=\columnwidth]{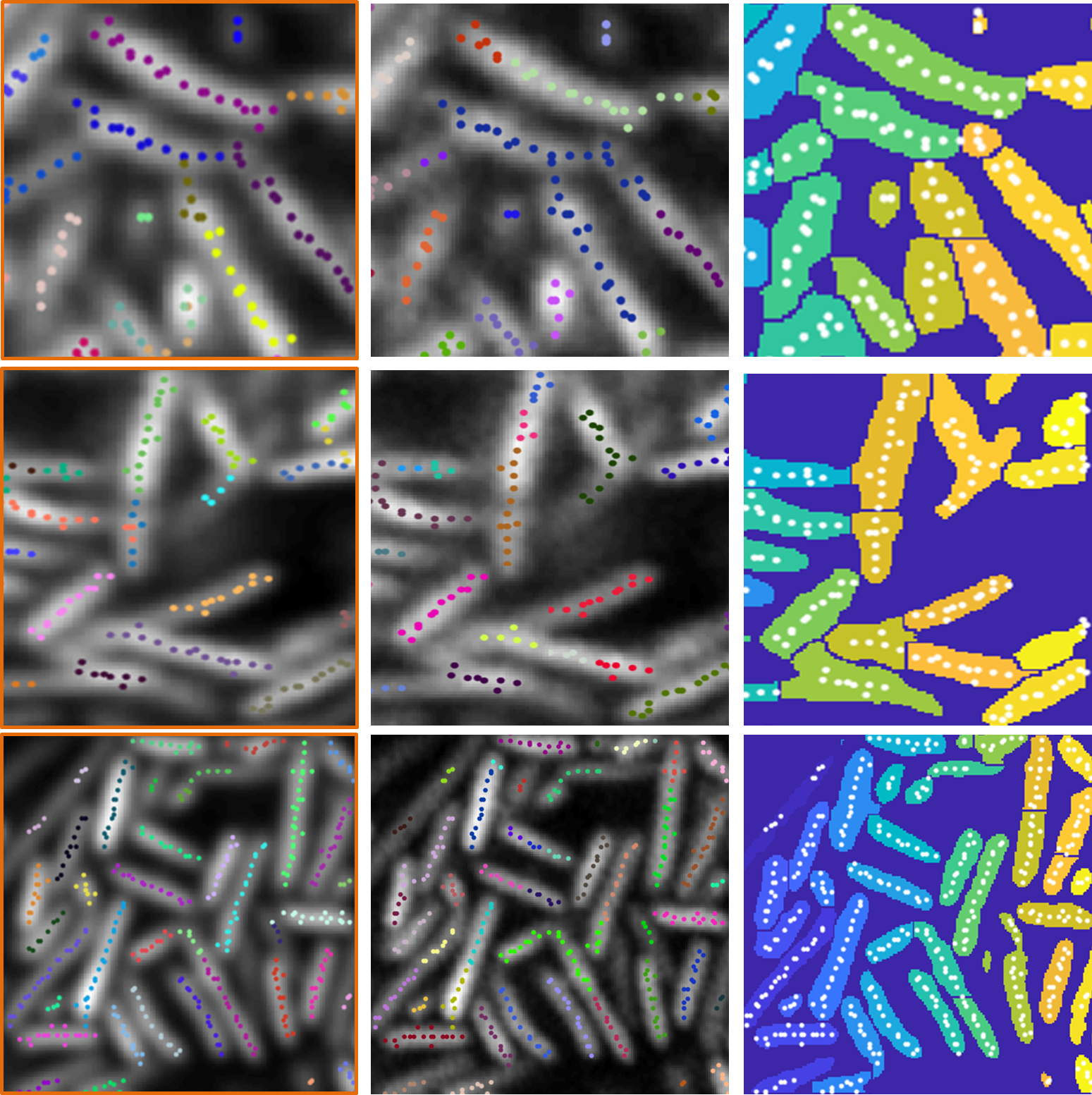}
    \vspace{-0.3cm}
	\caption{\small{Qualitative comparison for proposed method (first column) with DensityClust \cite{rodriguez2014clustering} (second column) and Single Cell Tracking \cite{yan2016vibrio} (third column).  For $\mathcal{L}$Cuts and DensityClust, different groups are marked with different colors and shown on the original image. The results of Single Cell Tracking are shown by overlapping the point cloud data on the segmented image, where different colors represent different single cell groups.}}
	\label{fig:compare}
	\vspace{-0.5cm}
\end{figure}

The performance of $\mathcal{L}$Cuts is analyzed qualitatively and quantitatively by comparing with two current methods used in the bioimaging community, \textit{DensityClust} \cite{rodriguez2014clustering} and \textit{Single Cell tracking} \cite{yan2016vibrio}.  %\textit{DensityClust} is a highly-rated clustering method that has favorable performance in classifying elements into different groups in different disciplines based on the relative density around each points. \textit{Single Cell tracking} is one of the recent and representative toolkit for reconstructing bacterial biofilms in the bio-chemisty field. It uses band-pass filter for image denoising and performs marker controlled watershed segmentation to separate single cells.
Based on the imaging technique and biological cell information, the parameter settings for $\mathcal{L}$Cuts are $sizeLimit = 60\text{ pixels}$, $distLimit = 5\text{ pixels}$ (maximum distance between nodes that have neighborhood), and $eccLimit = 0.9$.
The parameters are also tuned in the other two algorithms to achieve optimal performance in each case. In \textit{DensityClust}, we chose "Gaussian" mode for computing densities. Due to the manual input for the selection of cluster centers, we performed five times for each data and chose the best performance from all. In \textit{Single Cell tracking}, the watershed value is the key to optimizing the algorithm, where a value of one is used. Qualitative comparison is shown in \textbf{Fig. \ref{fig:compare}}. 

Two measures, grouping accuracy (GAcc) and counting accuracy (CAcc), are computed for quantitative comparison using $Dice$ = 2\text{TP}/(2\text{TP} + \text{FP} +\text{FN}), where $TP$= true positive, $FP$ = false positive, and $FN$ = false negative. GAcc accounts for the performance of how many nodes are correctly classified in each group (cell); while CAcc indicates the classification accuracy in terms of matching the final clusters with individual cells in the image. Here, individual cell regions are manually labeled as ground truth in the comparison.
%For both measures, the true positives (TP) are the number of nodes/cells that are correctly classified within each cell region. Here, cell regions are manually labeled as binarized ground truth as consistent standard in the comparison. For GAcc, false positives (FP) count the nodes in one group but they do not belong to the single cell that most of the group overlap with; while, false negatives (FN) represent those nodes missing in corresponding single cell. Similarly, for CAcc, FP is the counting for cells that are over-segmented and FN count for cells that have no corresponding cluster. 
\begin{table}[H]
\vspace{-0.2cm}
\begin{tabular}{c|cc|cc|cc}
     & \multicolumn{2}{c|}{$\mathcal{L}$Cuts}                 & \multicolumn{2}{c|}{DensityClust} & \multicolumn{2}{c}{SCT} \\ \hline

 &GAcc            & CAcc           & GAcc         & CAcc     & GAcc             & CAcc      \\ \hline
Best   & \textbf{95.9}       & \textbf{95.1}       & 94.6              & 92.3           & 94.4               & 94.1              \\
Worst    & \textbf{87.8}       & \textbf{85.2}       & 78.3              & 83.7             & 77.8                & 53.5              \\\hline
Avg & {\ul \textbf{91.6}} & {\ul \textbf{91.2}} & 85.9              & 87.2             & 87.7               & 86.5              
\end{tabular}
\vspace{-0.3cm}
\caption[position=bottom]{Quantitative comparison of $\mathcal{L}$Cuts with DensityClust \cite{rodriguez2014clustering} and Single Cell Tracking \cite{yan2016vibrio} using \textit{Dice} scores.}
\label{tbl:compare}
\vspace{-0.3cm}
\end{table}

%By comparison, \textit{DensityClust} and \textit{Single Cell Tracking} all present decent results in the 2D Airyscan images; however, they have their weaknesses. \textit{DensityClust} can efficiently find the cluster centers and classify the points into different groups, but it is hard to distinguish closely-connected components which have similar relative density. Another drawback of  \textit{DensityClust} is that the user needs to manually select the density centers, which can heavily affect the final output. Whereas, \textit{Single Cell Tracking} automatically detect the changes of cell morphology, but it is too sensitive when the cells have variant inter- and intra-cell intensities. It is also difficult to separate those touching components where bare changes are detected.

Overall, $\mathcal{L}$Cuts outperforms DensityClust and SCT in GAcc and CAcc by a margin of at least $4$\% on average. There are circumstances that some cells are misclassified in $\mathcal{L}$Cuts. One cause is the non-linearity of auto-produced point cloud data, especially when cells are randomly floating in the three-dimensional space. Another cause is the trade-off between the tolerance in distance/intensity changes and the continuity of the linear structure. 
%, in which case, prior information of the cell dimension is important in this application.
%Compare to the other 3D bacterial biofilm segmentation toolkits, $\mathcal{L}$Cuts overcomes the limitations of variant intensities inside of the densely packed biofilm and present steady work without the disturbance from the aforementioned factors in the 2D case.

$\mathcal{L}$Cuts can be directly applied on three-dimensional data. A preliminary result is shown in \textbf{Fig. \ref{fig:3dres}} with a Counting Accuracy of $97\%$. The point cloud data was generated by biofilm researchers in Gahlmann Lab. They manually labeled the centers of each bacteria slice by slice from $x$, $y$ and $z$ directions in Lattice Lightsheet microscopic image. The ground truth was manually grouped which reflects the actual single bacterium layout in 3D space.
\vspace{-0.3cm}
\section{Conclusion}
\label{sec:conc}
\vspace{-0.3cm}
We presented $\mathcal{L}$Cuts, a graph-based solution for finding linear structures in multi-dimensional spaces. $\mathcal{L}$Cuts outperforms the existing methods in majority of cases. Furthermore, $\mathcal{L}$Cuts enables automated processing of 2D and 3D images to identify individual bacteria in biofilms independent of the number of bacteria present. $\mathcal{L}$Cuts provides quantifiable information in the form of cellular positions, orientations, and the physical contact points between them. Beyond bacterial biofilms, $\mathcal{L}$Cuts can be extended to other biological applications in which boundaries are elusive but ridgelines of objects are accessible.
\newline

\bibliographystyle{IEEEbib}
\bibliography{lcuts}

\end{document}